\documentclass[aps,pre,twocolumn,floatfix,superscriptaddress,preprintnumbers,showpacs,showkeys]{revtex4}

\usepackage[english]{babel}
\usepackage{graphicx}
\usepackage{color}
\usepackage{amsmath,amsfonts}
\usepackage{subfigure}

\newcommand{\beq}{\begin{eqnarray}}
\newcommand{\eeq}{\end{eqnarray}}

\newcommand{\ev}[1]{\big\langle #1 \big\rangle}

\begin{document}
\bibliographystyle{h-physrev5}

\title{A numerical method for determining the interface free energy} 

\author{A. Hietanen}
\affiliation{College of Science, Swansea University,
  Singleton Park, Swansea SA2 8PP, UK}

\author{B. Lucini}
\affiliation{College of Science, Swansea University,
  Singleton Park, Swansea SA2 8PP, UK}

\begin{abstract}
We propose a general method (based on the Wang-Landau algorithm)
to compute numerically free energies that are
obtained from the logarithm of the ratio of suitable partition functions. 
As an application, we determine with high accuracy the order-order interface tension
of the four-state Potts model in three dimensions on
cubic lattices of linear extension up to $L=56$. The
infinite volume interface tension is then extracted at each $\beta$
from a fit of the finite volume interface tension to a known universal
behavior. A comparison of the order-order and order-disorder
interface tension at $\beta_c$ provides a clear numerical evidence of
perfect wetting.
\end{abstract}

%\preprint{CERN-PH-TH/2011-016}

%
\maketitle
 
\section{Introduction}
Interfaces play a major role in various physical
phenomena in, e.g., Statistical Mechanics, Soft Condensed Matter,
Particle Physics and Biology. Of particular interest are {\em fine}
(or {\em rough}) interfaces, whose long range fluctuations are determined by massless
modes. For those systems, the infrared properties are universal and
can be described by general models like the capillary wave model~\cite{Privman:1992zv} or
the Nambu-Goto string~\cite{Goto:1971ce,Nambu:1974zg}.

An interesting class of interfaces are those related to
the free energy of topological excitations. This is the case, e.g., for
the order-order interface in the Potts model or for the tension of the
't Hooft loop in SU($N$) gauge theories. The free energy of the
topological object for a system on a (hyper)cubic domain of linear
size $L$ can be extracted from the ratio of the 
partition function of the system in the presence of the topological
excitation (which can be enforced using suitable boundary conditions)
over the partition function of the system with periodic boundary
conditions. In more detail, if $\widetilde{Z}(L)$ is the partition
function in the presence of a topological excitation and $Z(L)$ the
partition function with periodic boundary conditions, $F_I$ , the free energy
of the interface (which we assume to be translationally invariant in
one direction) is given by
\beq
\label{eq:fi}
F_I(L) = - \log\frac{\widetilde{Z}(L)}{Z(L)} + \log(L) \ .
\eeq
The interface tension $\sigma$ is
then obtained as
\beq
\label{eq:sigmaas}
\sigma = \lim_{L \to \infty} \frac{F_I(L)}{L^{D-1}} \ ,
\eeq
with $D$ the dimensionality of the system. 

With some noticeable exceptions, it is not known how to determine from
first principles the analytical behavior of interfaces as a function of the couplings of
the system. Moreover, it is a
notoriously hard problem to access directly partition functions in
Monte Carlo simulations, since these quantities have exponential
fluctuations in the volume (see
e.g.~\cite{deForcrand:2000fi,deForcrand:2005rg}). Most
of the solutions adopted in the literature (e.g.~\cite{Kajantie:1988hn,Kajantie:1989iy,KorthalsAltes:1996xp}) 
consist in relating the interface tension (or its derivative) to
quantities that can be reliably determined via Monte Carlo
simulations. However, these methods generally introduce large
systematic and/or statistical errors. Hence, a direct determination of
the interface free energy as ratio of partition functions is
desirable.

\section{The Model}
The partition function $Z$ of a system at a temperature $T = 1/(k_B
\beta)$, with $k_B$ the Boltzmann constant, is obtained as the
integral (or the sum for discrete levels) over the energy $E$ of the
density of states $g(E)$ weighted with the Boltzmann factor $e^{-\beta
  E}$: 
\beq
Z = \int g(E) e^{- \beta E} \mbox{d} E \ . 
\eeq
Monte Carlo methods sample efficiently the distribution $g(E) e^{-
  \beta E}$, and are best suited for determining statistical averages
of observables with Gaussian fluctuations. An independent strategy for
studying statistical properties of a system consists in the
numerical determination of $g(E)$. This can be achieved using the
Wang-Landau algorithm~\cite{Wang:2001ab}. In this article, we 
propose a method to extract the interface tension using
Eqs.~(\ref{eq:fi})~and~(\ref{eq:sigmaas}) based on the Wang-Landau
algorithm. The method is tested on the four-state Potts model in three
dimensions.

In the $q$-state Potts model the fundamental
degrees of freedom are spin variables that can take the integer  values $0,
\dots, q-1$. The Hamiltonian of the model computed on a configuration
$\hat{q}$ is given by 
\begin{equation}
  H(\hat{q}) = 2 J \sum_{\langle ij\rangle}\left(\frac1q-\delta_{q_i,q_j}\right),
\end{equation}
where $J$ is the strength of the interaction, $\delta_{q_i,q_j}$ is the Kronecker
delta function of the spin variables $q_i,q_j$ on neighbor sites $i,j$ and the
sum $\langle ij \rangle$ is over nearest neighbors.
For a system of finite size, periodic boundary conditions in all directions
are imposed. The partition function is then given by 
\begin{equation}
\label{eq:zeta0}
Z = \sum_{\{\hat{q}\}} e^{- \beta H(q)} = \sum_E g(E) e^{- \beta E} \ ,
\end{equation}
where the first sum is over all possible configurations $\hat{q}$ and 
the second over all allowed energies (from now on, we redefine $\beta$ as
$J/(k_B T)$). At zero temperature, there
are $q$ stable vacua . In two and three spatial
dimensions, the system transitions from the low temperature ordered
phase, in which the spins are predominantly in one of the $q$ values,
to the high temperature disordered phase at the critical temperature
$T_c$.

For simplicity, we now specialize to the three-dimensional case. At
zero temperature, it is possible to enforce an interface separating 
two regions with two different vacua by imposing twisted boundary
conditions in one direction (e.g. the third direction). We consider
here only twists of one unit, i.e. twists for which spins $q_i$ at
points $i$ with coordinates $(x_1,x_2,L)$ are replaced by 
$(q_j + 1)$ mod $q$, where $j = (x_1,x_2,0)$ . We call
$\widetilde{H}(\hat{q})$ the corresponding Hamiltonian. 
The configuration that minimizes the
energy has a misalignement of the spins by one unit on a plane orthogonal to the third
direction. At finite temperature, this rigid interface between the two
vacua can fluctuate and near the phase transition becomes dominated by
massless modes (rough phase).  The partition function for the system
with an interface is given by  
\begin{equation}
\label{eq:zeta1}
\widetilde{Z} = \sum_{\{\hat{q}\}} e^{- \beta \widetilde{H}(q)} = \sum_{\widetilde{E}} \widetilde{g}(\widetilde{E}) e^{- \beta \widetilde{E}} \ ,
\end{equation}
where the tilde indicates that those quantities have to be computed
for the system with Hamiltonian $\widetilde{H}$.
With these definitions, for a system on a
cubic lattice of size $L$, the free energy and
the tension of the interface between two ordered states (order-order
interface) are given respectively by Eq.~(\ref{eq:fi})
and~Eq.~(\ref{eq:sigmaas}). 

\section{The numerical density of states}
To access directly the partition
functions~(\ref{eq:zeta0})~and~(\ref{eq:zeta1}),
we use the Wang-Landau algorithm~\cite{Wang:2001ab}.
This algorithm modifies directly the density of states by
performing a random walk in energy space. A
random update of a spin is accepted with a probability
$\min\left\{1,g(E)/g(E')\right\}$, where $E$ and $E^{\prime}$ are
respectively the energies before and after the update.
After the update, $g(E)$ is modified,
s.t. $g(E)\rightarrow kg(E)$, where now $E$ is the energy of
the configuration after the update. The update satisfies the detailed
balance in the limit $k \rightarrow 1$. However, starting from a $k \ne 1$
is important to obtain a first rough approximation of $g(E)$. For this
reason, we  begin the simulation with a relatively large value of $k$, namely
$k=3$. Then, when $g(E)$ has converged, we reduce $k \rightarrow
\sqrt{k}$ and repeat the cycle until $k$ is small enough for
the systematic errors to be significantly smaller than the statistical errors.
Each cycle defines one iteration of the algorithm.
We study cubic lattices of size $L=16, 20, 24, 28, 32, 40, 48,
56$. For each volume, we perform 20 independent simulations for both
periodic and twisted boundary conditions. For the smallest lattice, we start
from a constant density of states. For the other lattices, we use as an input
an interpolation of the density of states determined on the system with the
closest smaller size. After the simulation, $g(E)$ is normalised so that
$g(E_{\rm min})=\log(4)$ and $\widetilde{g}(\widetilde{E}_{\rm min}) =\log(4L)$.

In a Wang-Landau-type simulation, the convergence of the density of
the states and the saturation of the error present potential
issues~\cite{Yan:2003,Zhou:2005aa,Belardinelli:2007,Morozov:2007,Morozov:2009}.
The original implementation~\cite{Wang:2001ab} used a flatness criterion
for the histogram of the visits to the various energy levels: when the 
histogram is flat within some tolerance, we assume that $g(E)$ at that level of
iteration has converged. However, the tolerance is somewhat arbitrary. 
In~\cite{Zhou:2005aa}, it was proposed that the density of state converges
when each energy value is visited at least $1/\sqrt{\log k}$ times.
Increasing the number of visits will not decrease the statistical error. We will
refer to this proposal as the Zhou-Bhatt convergence criterion. This
criterion however does not address the convergence of the measured density
of state to the true density of state. In~\cite{Morozov:2007,Morozov:2009}
it was shown that possible systematic errors due to the convergence to
the wrong density of states can be eliminated if we require a number of visits
at least equal to $1/\log k$ for each energy level (Morozov-Lin convergence
criterion).

\begin{figure*}
  \begin{center}
    \includegraphics[width=0.95\textwidth]{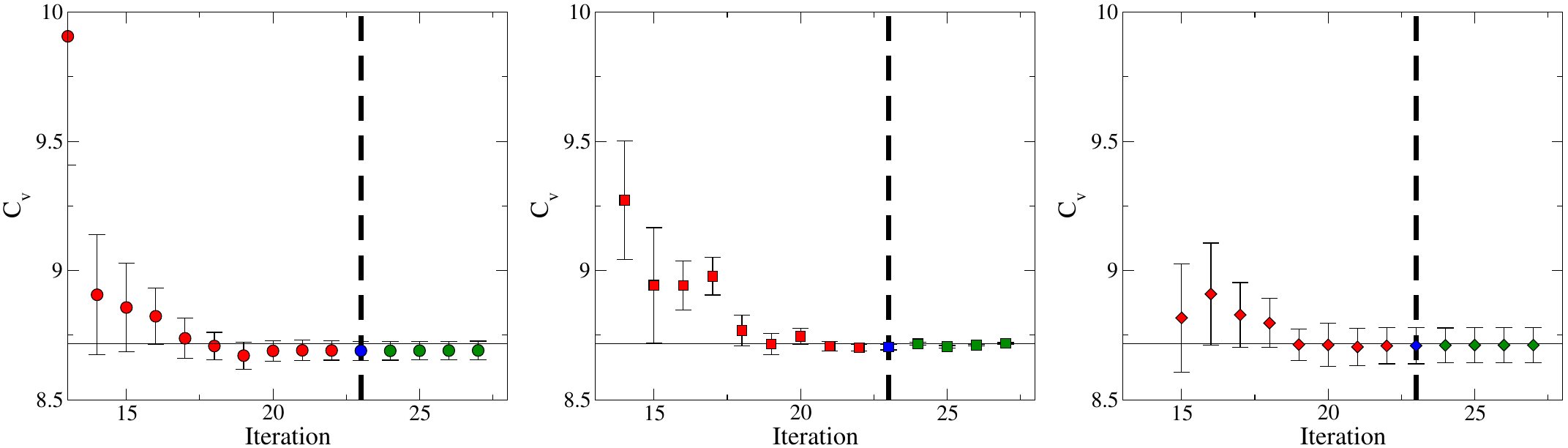}
    \caption{(Color
  online) Comparison between three different criteria to assess the
convergence of the density of states, for the specific heat $C_V$ at 
$\beta = 0.316$. The horizontal line is the central
value obtained with the Morozov-Lin criterion (middle plot) at the 27th
iteration. The left plot has been obtained with the Wang-Landau flat
histogram method, while the right plot shows the data obtained with the
Zhou-Bhatt criterion.
The vertical line marks the 23rd iteration.} \label{fig:convergence_cv}
  \end{center}
\end{figure*}
\begin{figure}
  \begin{center}
    \includegraphics[width=0.45\textwidth]{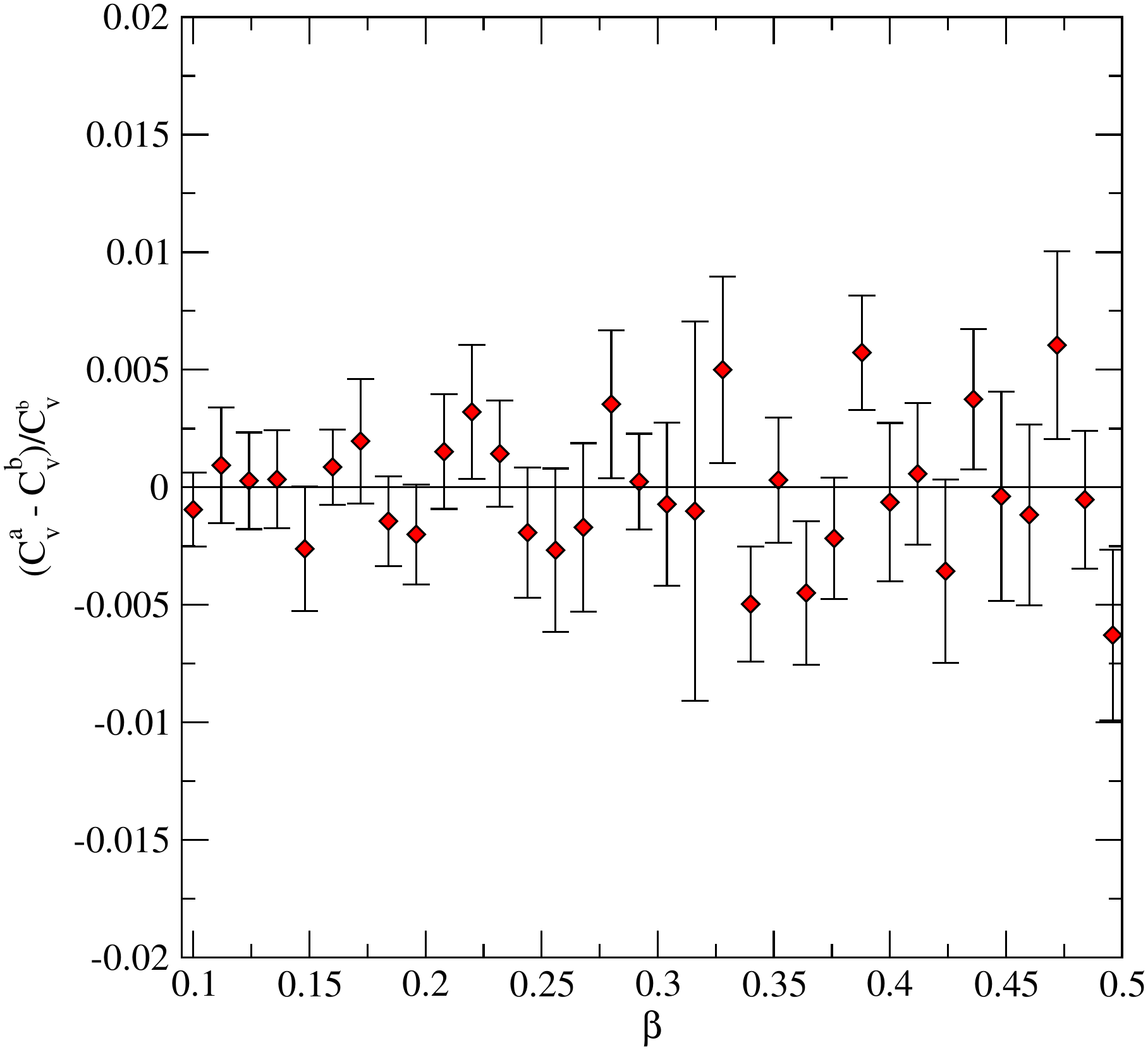}
    \caption{(Color online)
The difference between the specific heat determined with the
Zhou-Bhatt ($C_V^a$) and the Morozov-Lin criterion ($C_V^b$) divided
by $C_V^b$ (both taken after 23 iterations), on a wide range of $\beta$ for
the $16^3$ lattice.} \label{fig:cv_diff}
  \end{center}
\end{figure}

In Fig.~\ref{fig:convergence_cv},
we provide a comparison of the flat histogram criterion,
the criterion proposed in~\cite{Zhou:2005aa} and the criterion introduced
in~\cite{Morozov:2007} for the specific heat
\begin{equation}
  C = \frac{\beta^2}{L^3}\left(\ev{E^2}-\ev{E}^2\right) 
\label{cv}
\end{equation}
at $\beta = 0.316$ on a $16^3$ lattice. 
The figure shows that after 20 iterations all the three criteria are at
convergence. Moreover, the three estimates for the specific heat are the same
within errors, the typical size of the errors being a few percent.
This general feature is independent from $\beta$. An explicit comparison
of the Zhou-Bhatt criterion with the Morozov-Lin criterion focusing again on the
specific heat is provided in Fig.~\ref{fig:cv_diff} for
a wide range of $\beta$. This suggests
that for observables that are accurate to the level of the percent, among which
are interface tensions, the three criteria yield compatible results. In terms of
statistical errors, the Zhou-Bhatt criterion seems to yield the largest error
bars. However, the Zhou-Bhatt algorithm converges to the density of the states
in a CPU time that is a factor of 16 smaller than the original Wang-Landau and a
factor of 64 smaller than the Morozov-Lin criterion. This study suggests that
for our application the Zhou-Bhat criterion (i.e. number of visits
$\propto 1/\sqrt{\log k}$) is adequate from the numerical point of view
for the level of precision requested by our study. Hence, we used this
criterion to decide when a given iteration had converged. Based on the
study of Fig.~\ref{fig:convergence_cv}, we performed 23
iterations, which is a conservative estimate of the number of
iterations needed for the convergence of the algorithm.

\begin{figure}
  \begin{center}
    \includegraphics[width=0.45\textwidth]{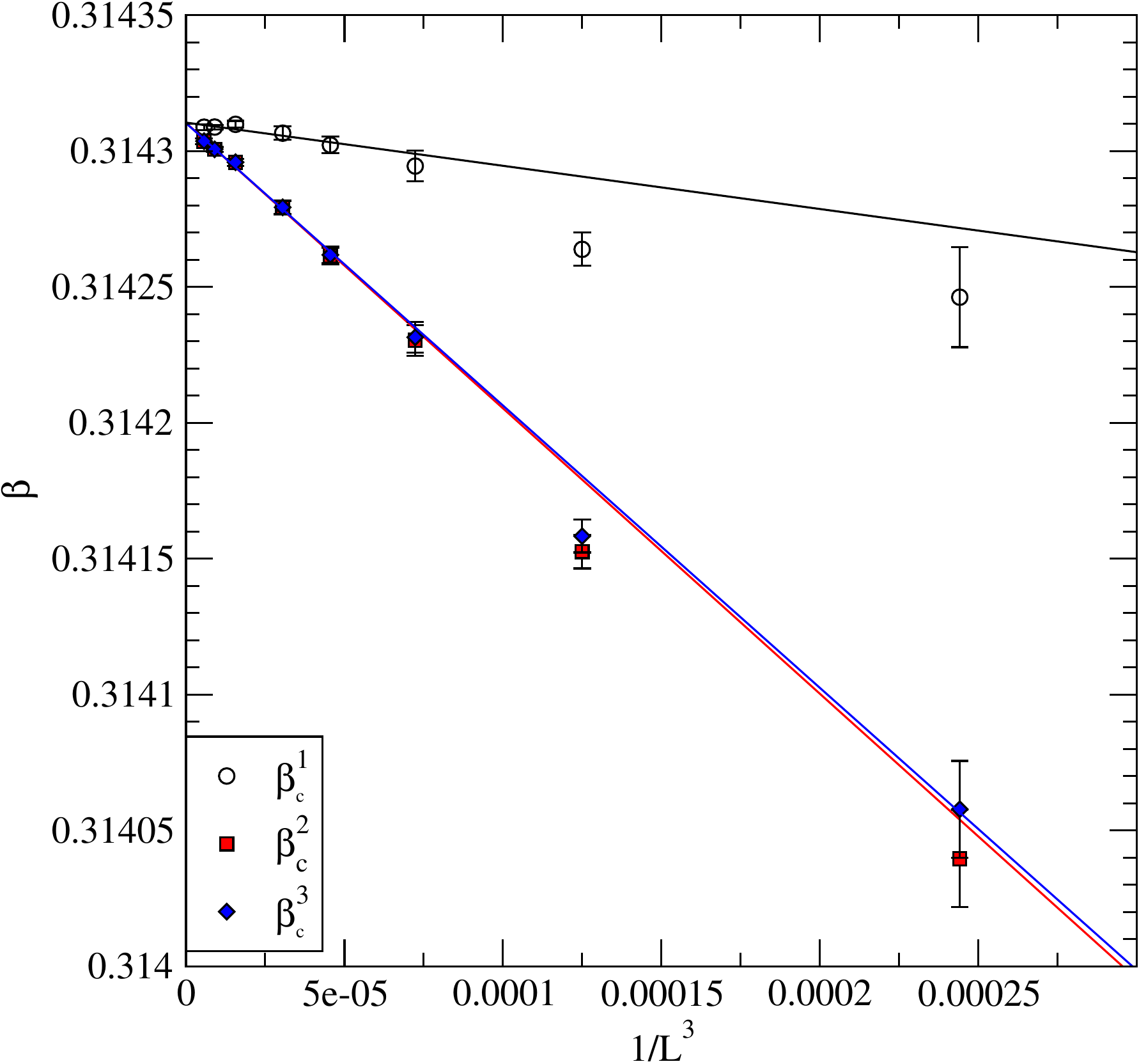}
    \caption{(Color online)
      Three different estimators for the critical temperature on
      finite lattices. \label{fig_tcext}}
  \end{center}
\end{figure}

\begin{figure}
  \begin{center}
    \includegraphics[width=0.45\textwidth]{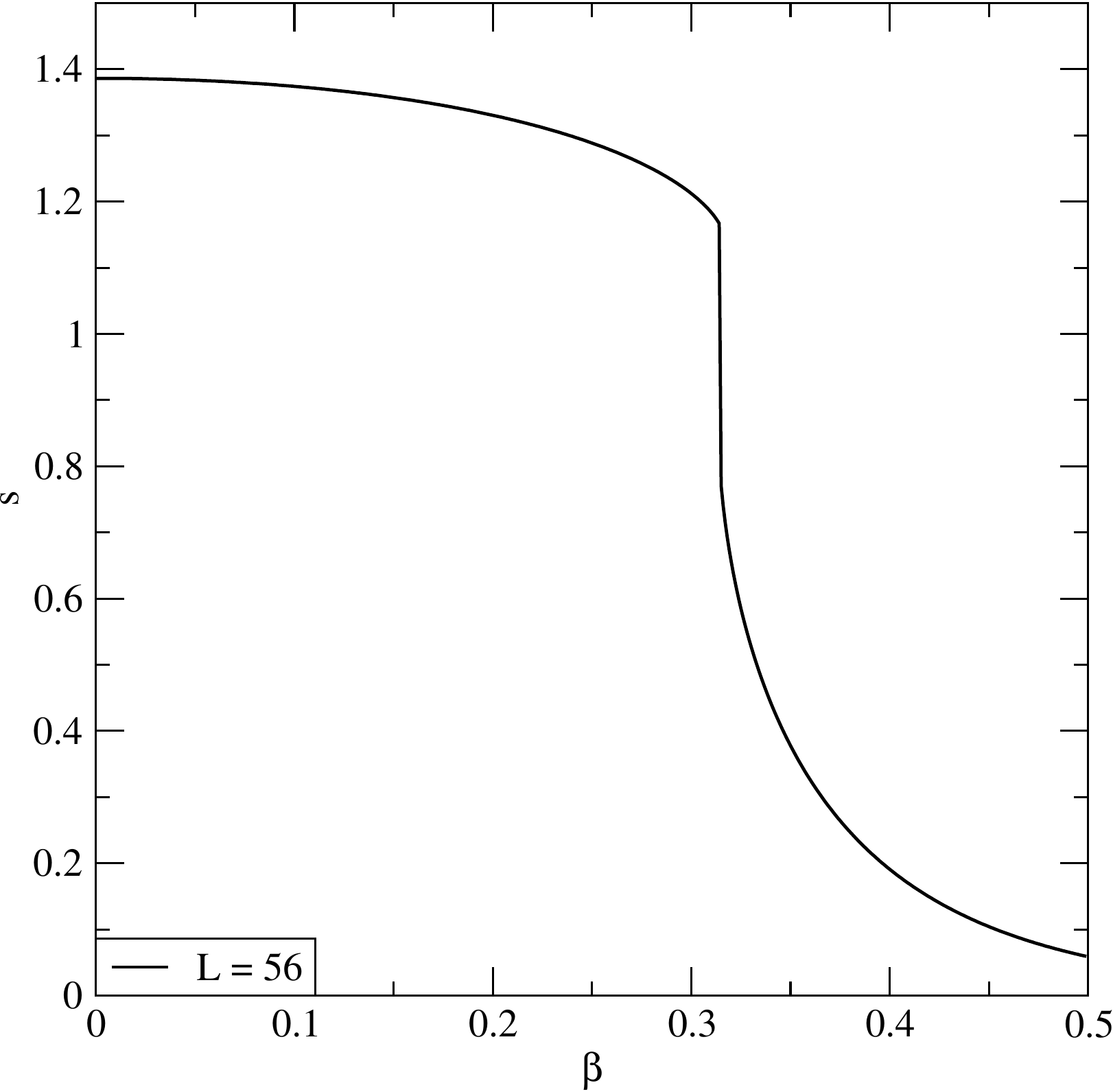}
     \caption{Entropy density as a function of $\beta$ for
       $L=56$. \label{fig_eqs}}
   \end{center}
\end{figure}

To perform further tests of our implementation,
we calculated some thermodynamical quantities like
the critical temperature, the latent heat and the entropy density, and compared
our results with Refs.~\cite{MartinMayor:2006gx,Bazavov:2008qg}.
Following~\cite{Bazavov:2008qg}, we use three different
estimators for the transition temperature on lattices with finite
extension. The first, $\beta^1_c$, is defined to be the
value for which the canonical distribution $P(E,\beta)=g(E)e^{-\beta E}$
has two equal maxima. The second ($\beta^2_c$) is the position of the central
energy of the latent heath, i.e. the value of $\beta$ satisfying
the equation
\begin{equation}
  e(\beta^2_c) = \frac12\left[e^{+}(\beta^2) + e^{-}(\beta^2)\right] \
  ,
\end{equation}
where $e^{\pm}$ ($e = E/N$, with $N=L^3$) are the locations of the maxima of the probability distribution at $\beta_c^2$.
The third ($\beta^3_c$) is the location of the maximum of the specific
heat~\ref{cv}. All the critical temperatures are extrapolated to the infinite
volume limit using the ansatz
\begin{equation}
  \beta^i_c(L) = \beta^i_c + \frac{c^i}{L^3} \ .
\end{equation}
We perform the extrapolation by using the six largest lattices. The
extrapolated values agree in the infinite-volume limit (see
Fig.~\ref{fig_tcext}). Averaging over the three determinations gives
$\beta_c = 0.3143103(9)$. 
The latent heat per site $\Delta e$ can be calculated from the maxima of the specific heat \cite{Challa:1986sk}:
\begin{equation}
  C_{\rm max}(L) = c + \frac14 (\beta_c)^2 (\Delta e)^2 L^3 \ .
\end{equation}
Our estimate is $\Delta e=1.16454(16)$. Finally, the numerically determined entropy density
\begin{equation}
  s= \beta(e-f) \ ,
\end{equation}
with $f = - \log Z/(\beta N)$ the free energy density, is plotted in Fig.~\ref{fig_eqs}.
All those quantities are always less than two standard deviations from the
corresponding determinations of~\cite{MartinMayor:2006gx}, which have been
obtained on larger lattices.

\section{Interface tensions}
\begin{figure}
  \begin{center}
    \includegraphics[width=0.45\textwidth]{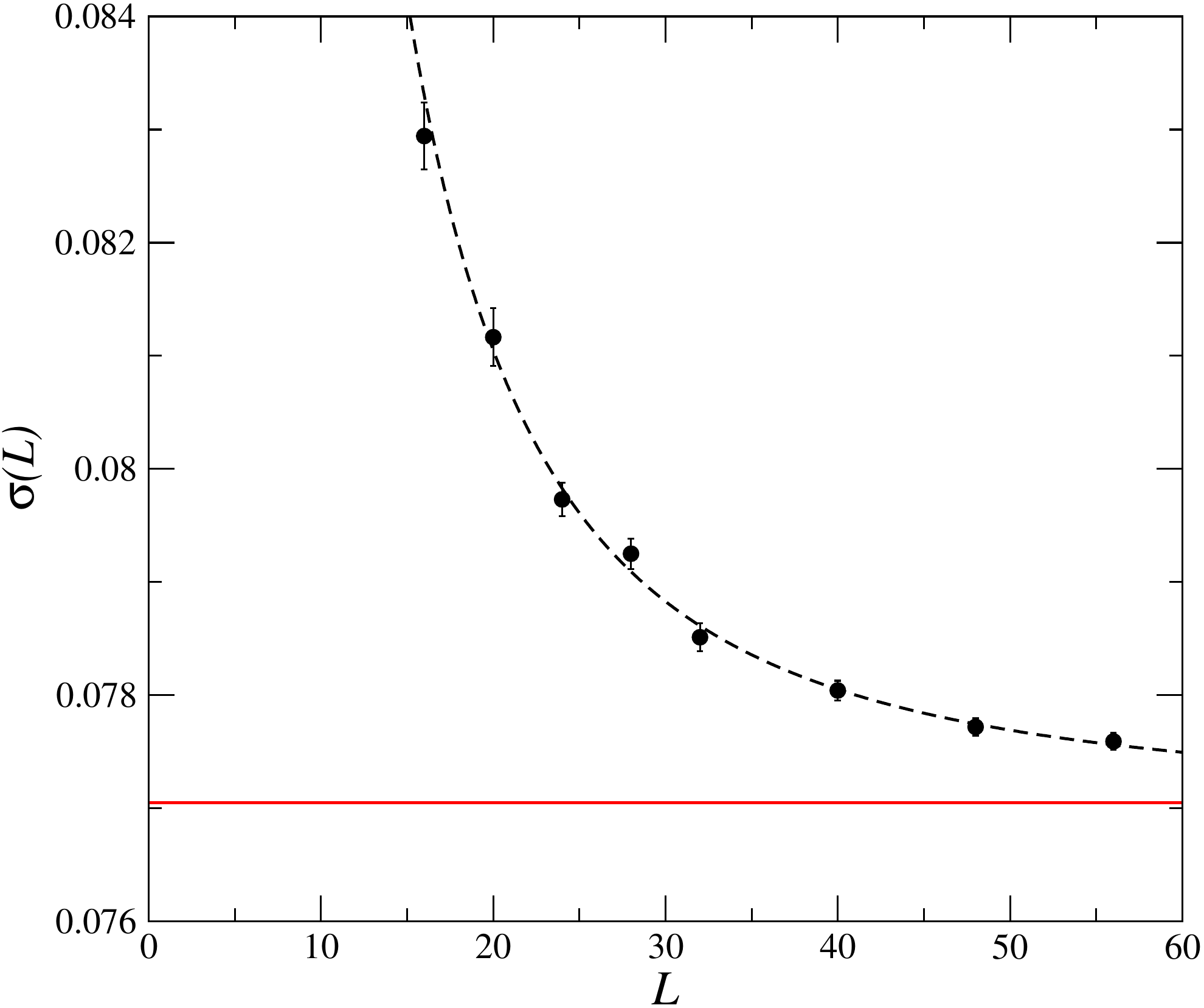}
    \caption{(Color online)
      The order-order interface tension $\sigma(L)$ at $\beta = 0.318$
      for different lattice sizes. The dashed line is a fit of the
      data according to Eq.~(\ref{eq:univinterface}), with fit parameters $\sigma_{oo}$, $c_2$ and $c_4$;
      the horizontal line indicates the extracted value for $\sigma_{oo}$.}
    \label{fig:univinterface}
  \end{center}
\end{figure}

\begin{figure}
  \begin{center}
   \includegraphics[width=0.45\textwidth]{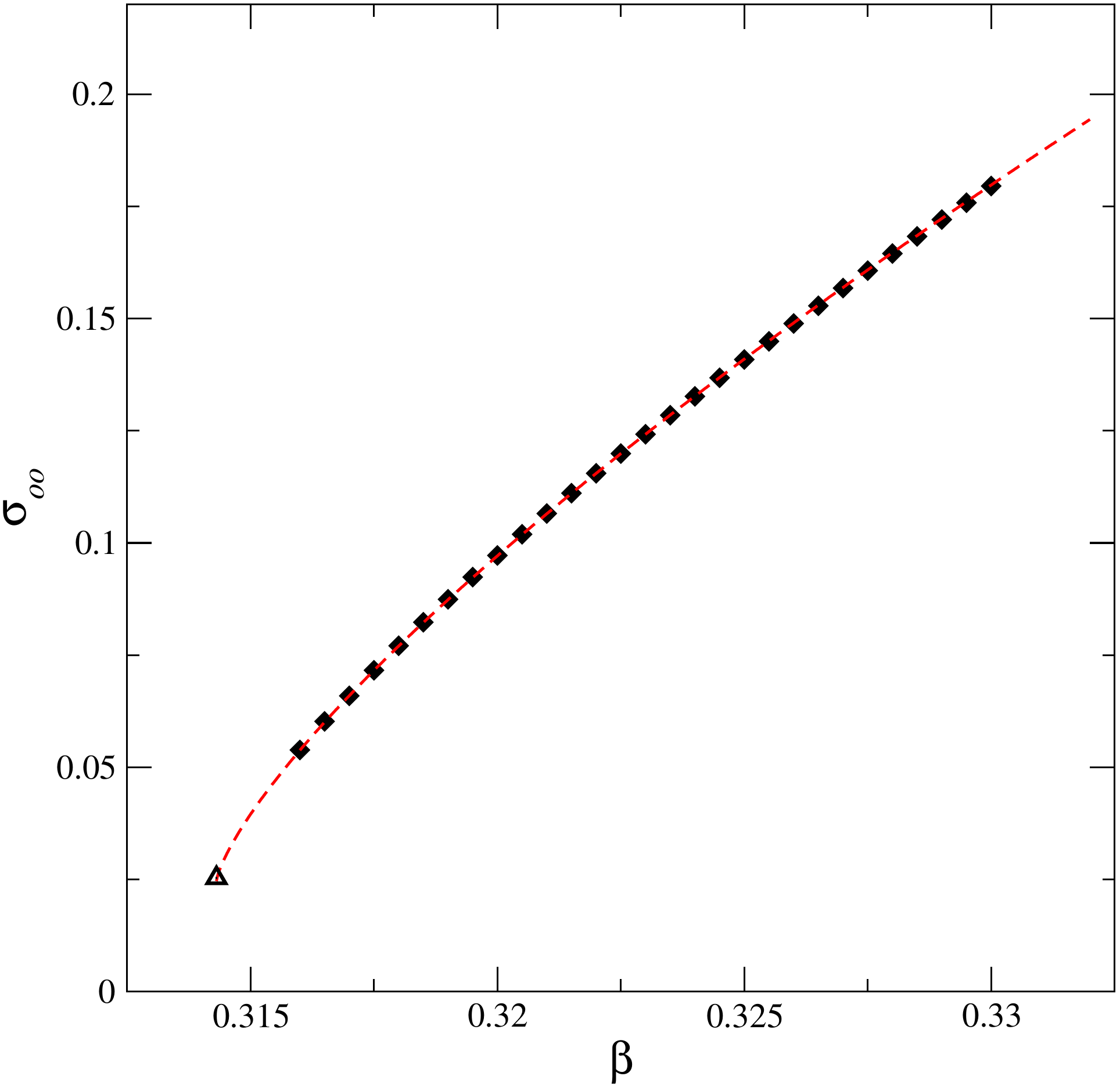}
    \caption{(Color online)
      The infinite volume order-order interface tension as a function of
      $\beta$. The fit (dashed line) is
      in excellent agreement with the hypothesis of perfect wetting
      (open triangle).}
    \label{fig:sigmaoo}
  \end{center}
\end{figure}

We now move to the discussion of our results for the interface
tension.
In the ordered phase, an interface can form
between two regions of space that are in two different vacua. This
interface is called the {\em order-oder interface}. 
Near the critical temperature, the dynamics of the order-order
interface $\sigma_{oo}$ is dominated by massless modes and its
infrared properties are universal (see
e.g.~\cite{Luscher:1980fr,Privman:1992zv}). The
asymptotic interface tension can then be extracted using the ansatz
(see e.g.~\cite{billo:2006zg,Caselle:2007yc})
\beq
\label{eq:univinterface}
\sigma_{oo}(L) = \sigma_{oo}  + \frac{c_2}{L^2} + \frac{c_4}{L^4} + \dots \ ,
\eeq 
where the order of the truncation of the expansion in $1/L^2$ is
determined by the accuracy of the data. As a consequence, a reliable
extraction of $\sigma$ requires accurate data.
From our simulations, we have extracted $F_I$ using
Eq.~(\ref{eq:fi}) and then the interface tension fitting the data
according to Eq.~(\ref{eq:univinterface}) truncated to ${\cal
O}(L^{-4})$. To reduce finite size effects, we included in
the fit only points for which $L \sqrt{\sigma} \ge 6$. Since the
ansatz~(\ref{eq:univinterface}) is expected to hold only at large
distances, our analysis was limited to values of $\beta$ for which
$\sigma_{oo}^{-1/2} \ge 3$. An example of the quality
of our data is given in Fig.~\ref{fig:univinterface}. 
The values of $\sigma_{oo}$ extracted from our fits are plotted in
Fig.~\ref{fig:sigmaoo}. The relative error on this quantity is at most
$3 \times 10^{-3}$, and is invisible on the scale of the figure. Near
$\beta_c$, the behavior of $\sigma_{oo}(\beta)$ can be parametrized
as
\beq
\sigma_{oo}(\beta) = \sigma_{oo}(\beta_c) + a\left( \beta - \beta_c
\right)^{\rho} \ . 
\eeq
Fitting this functional form to our results, we found that this
provides an excellent description of the data (a fit with 9 degrees of
freedom has $\chi^2/9 = 0.11$). We find $\sigma_{oo}(\beta_c) =
0.0249(6)$ and $\rho = 0.76(4)$. The quality of the fit is shown by
the dashed line in Fig.~\ref{fig:sigmaoo}.

At $\beta_c$, the {\em order-disorder interface}  separates a region in an ordered
state from one in the disordered state. The interface tension between
an ordered and the disordered phase, $\sigma_{od}$, can be
determined by looking at the probability distribution of the energy at
the critical temperature. In particular, if $P_{\rm max}$ is the peak
of the histogram when the two maxima have equal eight and $P_{\rm
  min}$ is the minimal height of the valley between the two peaks~\cite{Lee:1990ti,Berg:1992qua}, 
\begin{equation}
  2\sigma_{\rm od}(L) = \frac1{L^2}\log\left(\frac{ P_{\rm
        max}}{P_{\rm min}}\right) \ .
  \label{eq:ooestim}
\end{equation}
\begin{figure}
  \begin{center}
    \includegraphics[width=0.45\textwidth]{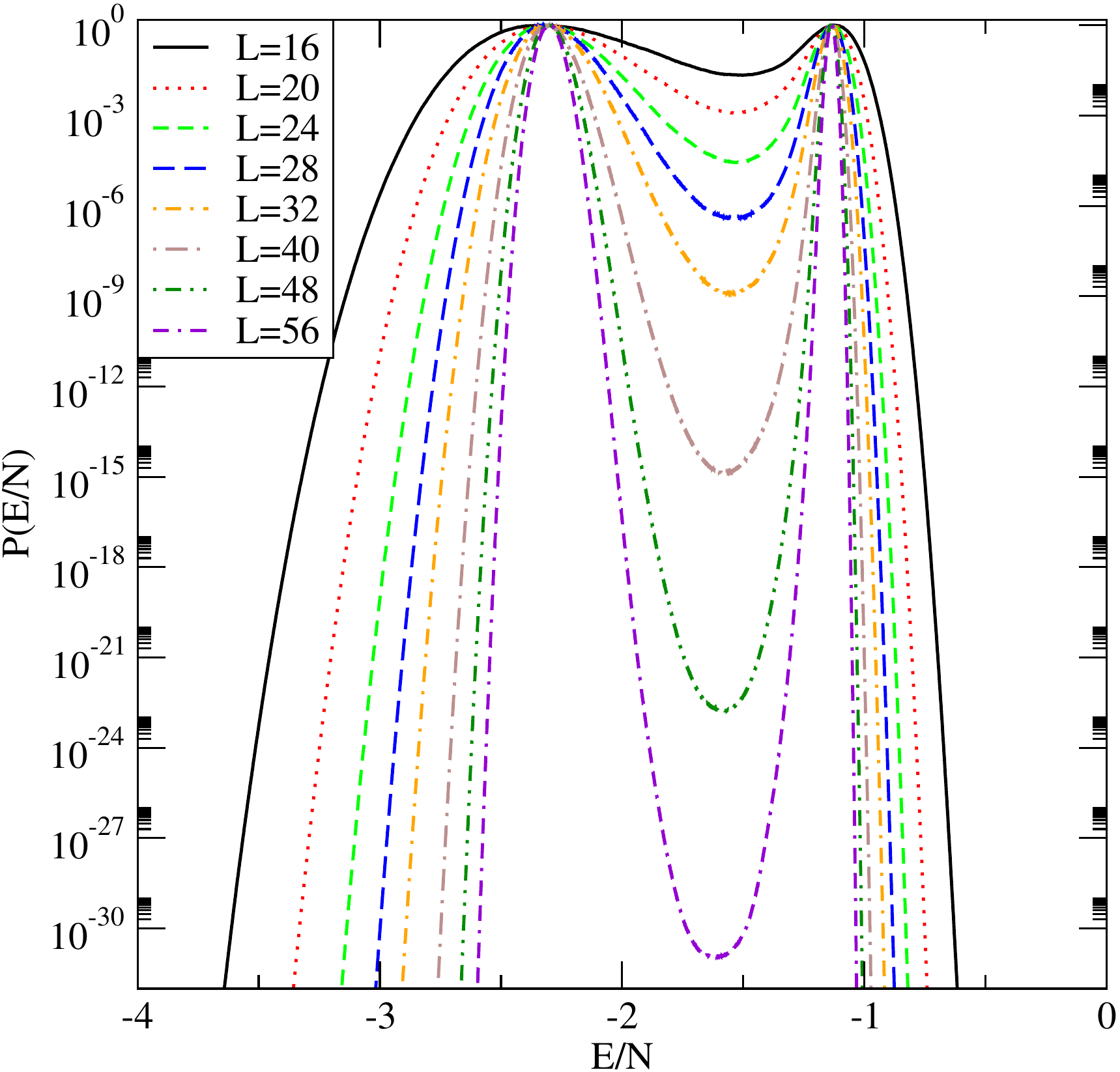}
    \caption{(Color online)
      The probability distribution at the critical
      temperature. The maxima are normalized to one.}
    \label{fig:peaks}
  \end{center}
\end{figure}
Our data for $P(E/N)$ as a function of $E/N$ are shown in Fig.~\ref{fig:peaks}. We use a fitting function of the form
\begin{equation}
  2\sigma_{\rm od}(L) = -\frac{\log L}{2 L^2} + 2\sigma_{\rm od}  +
  \frac{c_2}{L^2} + \frac{c_3}{L^3} + \frac{c_4}{L^4} \ .
  \label{dofit}
\end{equation}
Universality arguments~\cite{Luscher:2004ib,Aharony:2009gg}
suggest that $c_3 = 0$. Nevertheless, a correction at
this order is due to the fact that we use the finite-volume
estimator~(\ref{eq:ooestim}), as it can be shown with a simple saddle point
argument~\footnote{We thank V. Martin-Mayor for pointing out this
argument to us.}.
A fit of the data according to Eq.~(\ref{dofit}) performed excluding
the two smallest volumes yields $2\sigma_{\rm od} =
0.0252(4)$, with $c_3$ and $c_4$ both compatible with zero. Note that the
obtained value of $\sigma_{\rm od}$ is fully compatible with the perfect
wetting relationship
\beq
2\sigma_{\rm od} = \sigma_{\rm oo} \ , 
\eeq
which has been argued in~\cite{Borgs:1992qd} for the two-dimensional Potts
model. Our result is a clear indication that perfect wetting also holds for the 
Potts model in three dimensions. This is shown in Fig.~\ref{fig:sigmaoo}.

\section{Conclusions}
We have proposed a method for determining numerically
free energies
of interfaces or topological objects when the partition function with
given boundary conditions is required. The range of applicability
of our method includes not only statistical systems ($XY$ model,
Heisenberg ferromagnet etc.), but also gauge theories (e.g. the 't
Hooft loop tension in SU($N$) Yang-Mills). We have successfully tested
this method on the 3D four-state Potts model, for which we have
provided a very accurate determination of the order-order interface
below the critical temperature. This has enabled us to give clear
numerical evidence for perfect wetting in this model. 

\section*{Acknowledgements}
We thank M. Caselle and M. Panero for discussions and S. Kim and
A. Patella for comments on the manuscript. Correspondence with V. Martin-Mayor
is gratefully acknowledged.
The work of B.L. is supported by the Royal Society through the University 
Research Fellowship. The authors acknowledge support from STFC under
contract ST/G000506/1. The simulations discussed in this article have
been performed on a cluster partially funded by STFC and by the Royal Society.

\bibliography{potts}

\end{document}